\documentclass[useAMS,usenatbib]{mnras}
\usepackage{graphicx}
\usepackage[none]{hyphenat}
\usepackage{comment}
\usepackage[normalem]{ulem}
\usepackage{booktabs,caption,fixltx2e}
\usepackage[flushleft]{threeparttable}

\def\simgt{\lower.5ex\hbox{$\; \buildrel > \over \sim \;$}}

\title[G2 unlikely born in any known massive binary]{The Galactic Centre source G2 was unlikely born in any of the known massive binaries}
\author[D. Calder\'on et al.]{D. Calder\'on$^{1}$\thanks{E-mail:dcaldero@astro.puc.cl (DC)}, J. Cuadra$^1$, M. Schartmann$^{2,3}$,  A. Burkert$^{2,3}$, P. Plewa$^3$, \newauthor F. Eisenhauer$^3$, M. Habibi$^3$\\
$^1$Instituto de Astrof\'isica, Facultad de F\'isica, Pontificia Universidad Cat\'olica de Chile, 782-0436 Santiago, Chile\\
$^2$Universit\"atssternwarte der Ludwig-Maximilians-Universit\"at, Scheinerstr. 1, D-81679 M\"unchen, Germany\\
$^3$Max Planck Institute for Extraterrestrial Physics, P.O. Box 1312, Giessenbachstr. 1, D-85741 Garching, Germany
}

\begin{document}

\label{firstpage}

\date{Draft \today}

\pagerange{\pageref{firstpage}--\pageref{lastpage}} \pubyear{2018}
\maketitle

\def\simlt{\lower.5ex\hbox{$\; \buildrel < \over \sim \;$}}
\def\simgt{\lower.5ex\hbox{$\; \buildrel > \over \sim \;$}}

\begin{abstract}
	The source G2 has already completed its pericentre passage around Sgr~A*, the super-massive black hole in the centre of our Galaxy. 
	Although it has been monitored for 15 years, its astrophysical nature and origin still remain unknown.
	In this work, we aim to test the hypothesis of G2 being the result of a stellar wind collision.
	To do so, we study the motion and final fate of gas clumps formed as a result of collisions of stellar winds in massive binaries.   
	Our approach is based on a test-particle model in order to describe the trajectories of such clumps. 
	The model takes into account the gravitational field of Sgr~A*, the interaction of the clumps with the interstellar medium as well as their finite lifetimes. 
	Our analysis allows us to reject the hypothesis based on four arguments: 
	$i)$ if G2 has followed a purely Keplerian orbit since its formation, it cannot have been produced in any of the known massive binaries since their motions are not consistent; 
	$ii)$ in general, gas clumps are evaporated through thermal conduction on very short timescale ($<100\,\rm yr$) before getting close enough to Sgr~A*;  
	$iii)$ IRS~16SW, the best candidate for the origin of G2, cannot generate clumps as massive as G2; 
	and $iv)$ clumps ejected from IRS~16SW describe trajectories significantly different to the observed motion of G2. 
\end{abstract}


\section{Introduction}

	\cite{G12} discovered an enigmatic source traveling on a nearly radial orbit to the super-massive black hole (SMBH) of our Galaxy, Sgr~A*. 
    	Several groups have been monitoring the so-called G2 object since then, aiming to capture its interaction with Sgr~A* as well as to understand its astrophysical nature \citep{W14,V15,P17}. 
    	The possibility of observing a tidal disruption event in the Galactic Centre is a unique opportunity to investigate accretion physics and constrain the properties of the accretion flow. 
    	$L'$-band observations with aid of adaptive optics show G2 as a dusty, unresolved source before and after pericentre passage \citep{G12, W14}. 
    	On the other hand, \textit{SINFONI/VLT} Br$\gamma$ and \textsc{HeI} line observations revealed an extended source of ionised gas being affected by tidal shearing due to Sgr~A*'s strong gravitational pull \citep{G13a,G13b,P15}. 
    	The latest observations complete 15 years of monitoring its orbit, including its recent pericentre passage \citep{P17}.
    
    	Since its discovery there has been an interesting debate regarding G2's nature. 
    	The proposed scenarios can be broadly divided in two groups: the compact source and the purely gaseous cloud. 
    	The former refers to hypotheses that consider there is/was an object such as a star or a planet within G2. 
   	On the contrary, the latter refers to models in which G2 is a purely gaseous and dusty object.
    	Among compact source scenario explanations are an evaporating circumstellar disc \citep{MC12,ME12} or proto-planet \citep{M15}, a star with a gaseous and dusty envelope \citep{B13,D14,V15,B16}, and the result of the merger of a binary system \citep{W14,Pr15}. 
    	On the other hand, if G2 is solely made out of gas and dust it might have originated from the slow stellar wind ($300-600\rm\ km\ s^{-1}$) of a luminous blue variable \citep{B12}, as the result of a partial tidal disruption of a late-type giant star by the action of the SMBH \citep{G14}, or due to a recent nova outburst that ejected a ring-like shell of gas \citep{MM12}. 
	An additional explanation is that the source is part of a clumpy larger stream of gas formed in a stellar wind collision \citep{B12,G12,S12,C16}. 
    	This hypothesis is supported by the presence of another object, called G1, that shares similar characteristics with G2, including almost the same orbit but 13 years forward in time \citep{P15}. 
    	Moreover, the same study showed the presence of a structure of material trailing G2 on a slightly different orbit. 
    	Recently, \cite{P17} showed that this ``tail" of G2 is actually following G2. 
    	Tracing back its motion it seems to be coming from the massive binary IRS~16SW which suggests a possible origin. 
	Furthermore, the alignment of G2 and IRS~16SW orbits supports this idea \citep{G12,B12}. 
    	Another argument in favour of this scenario was shown in \cite{C16}. 
	In that study we analysed stellar wind collisions of all possible pairs of mass-losing stars in the Galactic Centre to see whether clump formation could take place or not. 
	There, we found that wind collisions of massive binary systems, and not of encounters between single stars, satisfied the requirements to generate such gas clumps.
	Specifically, we concluded that IRS~16SW is the best candidate to produce clumps based on its binary and stellar wind properties.

    	In this work, we aim to test whether a massive binary system, like IRS~16SW, can indeed create and place G2 on its observed orbit assuming a purely gaseous nature. 
    	The best way of testing this hypothesis would be to run a self-consistent hydrodynamical simulation of the wind-emitting binary orbiting Sgr~A*. 
    	However, the scales that need to be covered in space and time span at least 5 and 7 order of magnitudes, respectively, which makes the problem an overwhelming computational challenge. 
    	Therefore, in this study we opt for analysing the observed motion of G2 and study whether it is consistent with an origin in a massive binary. 
    	Furthermore, we attempt to reproduce G2's motion by clumps produced in IRS~16SW. 
    	To do so, we set up and run several test-particle simulations in order to quantify the probability of ejecting a clump that mimics the trajectory of G2. 
    
    	This paper is divided as follows: Section~\ref{sec:cwb} describes the clump formation process in colliding wind binaries. 
    	In Section~\ref{sec:g2clump}, we introduce the hypothesis of G2 being formed in IRS~16SW and discuss whether it is consistent with the current knowledge of these objects. 
    	Section~\ref{sec:ejec} introduces and describes our test-particle simulations of IRS~16SW ejecting clumps while orbiting Sgr~A*. 
    	In Section~\ref{sec:res}, we present the results of our simulations. 
    	Section~\ref{sec:disc} compares the results of our model with G2's motion and discusses the limitations of our model.
    	Finally, in Section~\ref{sec:conc} we close with some final remarks and present the conclusions of this work.

\section{Clump formation in colliding wind binaries}
\label{sec:cwb}
	A colliding wind binary is a system of two gravitationally bound stars, whose winds collide between them. 
	This collision creates a hot slab filled with shocked material reaching temperatures of $10^6-10^7\rm\ K$ in the case of Wolf-Rayet stars. 
	If the material can radiate its energy rapidly enough, the slab becomes thinner and denser. 
	If perturbed, the slab is prone to suffer \textit{thin-shell instabilities} \citep{V83, V94}. 
	These mechanisms are excited due to the misalignment between the wind ram-pressure and the thermal pressure within the slab. 
	Furthermore, the velocity difference between both winds can excite the Kelvin-Helmholtz instability which can act simultaneously with other instabilities. 
	However, \cite{L11} showed that the \textit{non-linear thin shell instability} \citep[hereafter NTSI][]{V94} is the one that dominates the long-term evolution of the slab due to its large-scale perturbations. 
	Therefore, we expect this mechanism to produce clumps which then are ejected to the ISM. 

	Recently, we conducted a study aiming to estimate analytically the clump masses formed through the NTSI in the Galactic Centre \citep{C16}. 
	We showed that Earth mass clumps can be produced for certain combinations of wind speed and stellar separation, for stars with strong outflows. 
	Moreover, given the known stellar population in the GC, we found that massive binary systems are the most promising clump sources within the inner parsec. 
	
	\subsection{Massive binaries orbiting Sgr~A*}
	
		The inner parsec of our Galaxy hosts about 30 Wolf-Rayet stars. 
        		Photometric and spectroscopic studies have provided valuable information of their orbits and stellar winds. 
        		Years of monitoring allowed to identify three binary systems among this sample \citep{M06,P14}.  
		However, there are other four sources considered as binary candidates \citep[see][]{P14}. 
		Although some of them showed changes on either brightness or radial velocity, there are not enough observations to confirm their binarity. 
		Furthermore, only two of them are within half a parsec from Sgr~A*. 
		The other two are at $\sim1.5\rm\ pc$, so latest surveys did not include them.
        		Therefore, throughout this work we focus uniquely on systems already confirmed as binaries.
        
		As we stated previously, there are three confirmed massive binaries inside the central parsec of the Galaxy. 
		The first one, IRS~16SW, was identified as an eclipsing binary with a period of 19.5-d by \cite{M06}.
		Its symmetric light curve indicates it is composed of equally massive stars. 
		Their mass was derived from its dynamics to be about $\sim50$~M$_{\odot}$ for each star. 
		The spectrum is consistent with the presence of strong outflows ($\sim10^{-5}\rm\ M_{\odot}\ yr^{-1}$) whose terminal velocity is $\sim600$~km~s$^{-1}$ \citep{M07,C08}. 
		More recently, \cite{P14} identified two new OB/WR binaries: IRS~16NE, a long-period Ofpe/WN9 binary with a period of 224-d, and E60, an eclipsing Wolf-Rayet binary with a period of 2.3-d. 
        	        	The most relevant properties of these massive binaries are summarised in Table~\ref{tab:binaries}.
        
\begin{figure}
			\centering
			\includegraphics[width=0.49\textwidth]{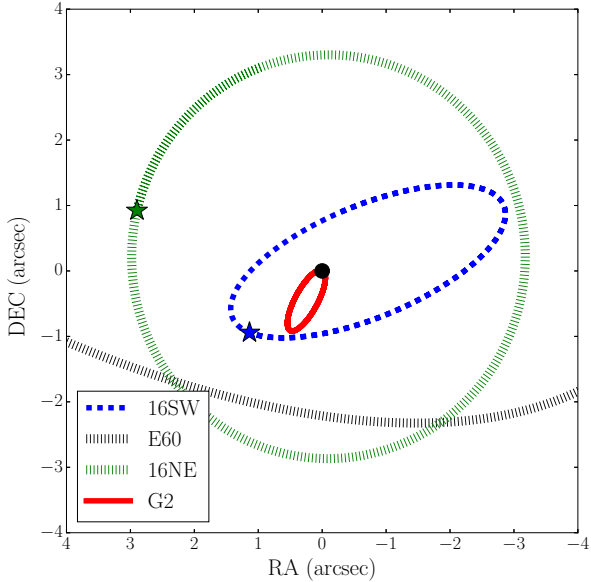}
			\caption{Sky-projected orbits of the massive binary systems and G2 around Sgr~A*.
			The dash-dotted black, dash-dotted green and dashed blue lines show the trajectories of IRS 16SW, IRS 16NE and E60, respectively. 
            		Star symbols stand for their current positions. 
            		E60's current position is outside of this region
			Notice that G2's apocentre roughly matches with the orbit of IRS~16SW.}
			\label{fig:orbits}
		\end{figure}

		In Figure~\ref{fig:orbits}, we show the sky-projected orbits of the known binaries around Sgr~A*. 
		Throughout this work we used the most recent orbital data available in the literature for the binaries as well as for G2 \citep{G17,P17}.
		 Nevertheless, it is important to remark that only IRS~16SW's orbit is completely determined (i.e., observed sky-positions, proper motions, radial velocity, and sky-acceleration).  
        		The orbits of the other two binaries have not been entirely constrained yet, as no acceleration has been detected so far.  
        		The orbits shown in the figure were chosen by iterating over the unknown position along the line of sight, and minimising the resulting orbital eccentricity.  
		Also, by the same procedure we have calculated lower limits for their pericentre distances (see Table~\ref{tab:binaries}). 
		Notice that the {\it actual} pericentre distance of IRS~16SW around Sgr~A* is shorter by $\approx 2/3$ than the {\it minimum possible} pericentre of either of the other binaries. 
		For them, the free-fall timescale is significantly longer than the typical clump lifetime. 
        		Then, it is not possible for clumps created in those binaries to get close to Sgr~A* (see Section~\ref{sec:mass}). 
        		Therefore, in this work we focus our study on clumps ejected from IRS~16SW. 
        
       		It is important to highlight that the apocentre of G2's orbit roughly coincides with IRS~16SW orbit (see Figure~\ref{fig:orbits}). 
		This fact also gives us a hint of the possible origin of G2 from this binary. 
        
        		\begin{table*}
		\begin{threeparttable}
        			\centering
            		\caption{Properties of massive binaries in the vicinity of Sgr~A*. }
            		\begin{tabular}{|l|c|c|c|c|c|c|c|}
            			\hline
                			Name	&	Type	&	$a\sin i$ $\rm (au)$	&	Eccentricity	&	Period $\rm (d)$	&	$\dot{M}_{\rm w}$ $(\rm M_{\odot}\ yr^{-1})$				&	$V_{\rm w}$ $(\rm km\ s^{-1})$			&	 $r_p$ $\rm (arcsec)$\\
				(1)	&	(2)	&	(3)		&	(4)			&	(5)		&	(6)					&	(7)				&	(8)\\
            			\hline
                			\hline
                			IRS~16SW	&	Ofpe/WN9	&	0.3	&	$\sim0$		&	19.5		&	$\sim 10^{-5}$		&	600	&	1.5 \\ 
                			IRS~16NE	&	Ofpe/WN9	&	1.2	&	$\sim0$		&	224		&	$\sim 2\times10^{-5}$	&	650	&	$>2.8$ \\ 
                			E60		&	WR binary	&	0.1	&	0.3			&	2.3		&	$\sim 5\times10^{-6}$	&	750	&	$>2.3$\\
                			\hline         
            		\end{tabular}
			\label{tab:binaries}
			\begin{tablenotes}
				\item \textit{Notes.} 
				Column~1: binary system ID. 
				Column~2: Spectral classification \citep[if available;][]{M07}. 
				Column~3: binary semi-major axis. 
				Column~4: binay eccentricity. 
				Column~5: binary period. 
				Binary orbital properties were taken from \citet{M06} and \citet{P14}.
				Column~6: stellar wind mass loss rate. 
				Column~7: stellar wind terminal velocity. 
				Stellar wind properties were obtained from \citet{M07} and \citet{C08}. 
				Column~8: pericentre distance of the binary system around Sgr~A* constrained from the observed sky-projected positions, proper motions and line-of-sight velocity \citep{P06,G17}.
			\end{tablenotes}
		\end{threeparttable}
        		\end{table*}
	
\section{G2 as a gas clump from IRS~16SW}
\label{sec:g2clump}   

	There are three pieces of evidence pointing to IRS~16SW as G2's origin. 
	Firstly, the cloud orbit is coplanar with the so-called clockwise disc, which includes the binary \citep{G12,P15}. 
	Secondly, G2's apocentre lies roughly on the orbit of IRS~16SW (see Figures~\ref{fig:orbits} and~\ref{fig:timing}). 
	Finally, the gas stream following G2 seems to have been originated in IRS~16SW \citep{P17}. 
	In this section, we study if this hypothesis is consistent with G2's motion and the expected lifetime of gas clumps in this region.
   
	\subsection{The Keplerian orbit}
   	\label{sec:kepler}
   
\begin{figure}
			\centering
			\includegraphics[width=0.45\textwidth]{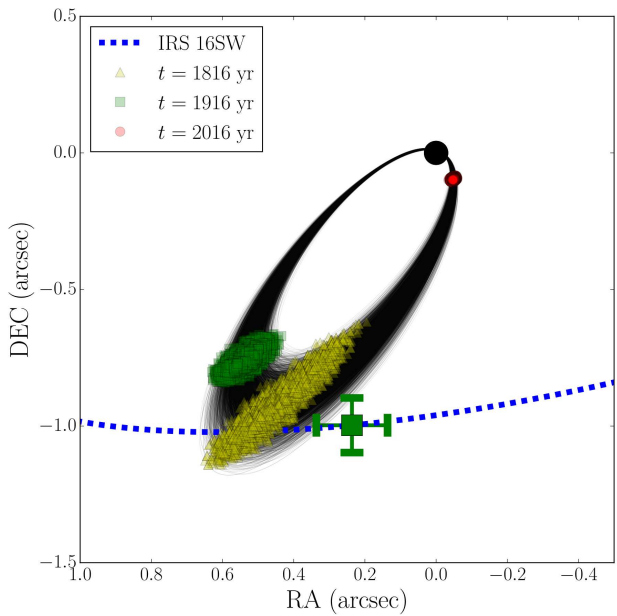}
			\caption{Sky-projection of the best fit orbit to G2's motion plotted with $3\sigma$ errors \citep{P17} shown as solid black lines. 
            		Part of the orbit of IRS~16SW is also shown as dashed blue line. 
            		Coloured symbols represent positions at different epochs: $t=1816\ yr$ (yellow triangles), $t=1916\rm\ yr$ (green squares), $t=2016$ (red circles). 
			The error on the projected orbit of IRS~16SW in $t=1916$ is $\sim0.1\rm\ arcsec$. 
			The big black dot at the origin represents Sgr~A*.
			Notice that G2's apocentre passage is overlaid on IRS~16SW's orbit within the errors.}
			\label{fig:timing}
		\end{figure}

   		Observations of G2's orbit currently span around 12 years, reaching its pericentre.  
		These data can be well described with a Keplerian orbit \citep{G12,G13a,G13b,P15,V15,P17}, i.e., within the error bars there is no apparent deviation from a ballistic motion. 
        		Let us consider G2 has moved on this orbit since it was formed.
     		This assumption allows us to draw conclusions from the analysis of G2 and IRS~16SW observed orbits. 
        		In Figure~\ref{fig:timing}, we show sky-projected possible orbits of G2 (solid black lines) and the binary (dotted blue line) around the SMBH. 
        		We highlighted positions at $t=1816,1916, 2016\rm\ yr$, the latter being the epoch of the latest G2 observations. 
		Notice that the positions of IRS~16SW and G2 do not coincide in any time shown. 
        		At a given epoch $t$ both sources are separated by at least $\sim0.25\rm\ arcsec\sim 2000\rm\ au$, 
		even taking into account the uncertainties associated to their infered orbits (see Figure~\ref{fig:timing})\footnote{The uncertainty on IRS 16SW's position in 1916 was calculated by simple error propagation from its currently measured position and velocity.  
		Given that the time interval considered is much shorter than its orbital period, no further analysis is needed}. 
        		Thus, G2 cannot have been created in IRS~16SW. 
     		However, we have to be careful with this conclusion.
		We have to bear in mind that in this analysis we have not considered systematic uncertainties on G2's orbit. 
        		As the source has been monitored solely around its pericentre passage the pericentre time and distance are well constrained. 
        		On the contrary, the apocentre is very uncertain. 
        		In the previous analysis we just extrapolated the observational data based on the Keplerian orbit assumption. 
        		Then, although this suggests the hypothesis is not feasible we cannot quantify the robustness of this conclusion with the available data.
        
	\subsection{Clump lifetimes}
	\label{sec:life}
    
    		A critical constraint to conceive G2 as a gas clump is the lifetime of such a clump.
       	 	Once formed, clumps are immediately ejected into the ISM that close to Sgr~A* is a very hostile environment. 
        		The region is dominated by shocked stellar winds blown by tens of Wolf-Rayet stars.
        		Therefore, the ISM is composed by very hot, diffuse plasma. 
        		The properties of the medium are very different compared to the ones of clumps.
        		Typically, the medium is at a temperature of $\sim10^7\rm\ K$ while clumps are kept at $\simgt10^4\rm\ K$  by the strong UV radiation from the young stars present in the region.
        		As we know, cold gas clumps embedded in a hot, diffuse medium can be seriously affected by thermal conduction.
		The rate of evaporation due to this process depends on the ratios of density and temperature between the clumps and the medium. 
        
		Based on \cite{C77} and \cite{B12}, the mass-loss rate due to thermal conduction of a cloud with a given mass $m_{\rm c}$ and distance from Sgr~A* $r$ can be estimated through the expression 
		
		\begin{eqnarray}
		\label{eq:mdot}
			\dot{M}_{\rm evap}	&	=	&	1.25\times10^{-2}{\rm\ M_{\oplus}\ yr^{-1}}\nonumber\\
							&		&	\times(\alpha+1)^{\frac{1}{6}}\left(\frac{n_0}{100\rm\ cm^{-3}}\right)^{\frac{1}{3}}\nonumber\\
            						&		&	\times\left(\frac{m_{\rm c}}{1{\rm M_{\oplus}}}\right)^{\frac{2}{3}}\left(\frac{1.7\times10^{17}\rm\ cm}{r}\right)^{\frac{2\alpha-1}{6}},
		\end{eqnarray}
		
       	 	\noindent where $n_0$ and $\alpha$ are the normalization and power-law of the medium density profile, respectively; i.e., $n(r)=n_0(r_0/r)^{\alpha}$, which were chosen to reproduce the X-ray observations of the region (see Appendix~\ref{sec:ism}).
        		Notice that this expression was derived in the saturation limit, therefore, it is not very sensitive to the presence of magnetic fields \citep{C77}.
		
		Then, the evaporation timescale is
	 
		\begin{eqnarray}
		\label{eq:evap}
			\tau_{\rm evap}	&	\approx 	&	\frac{80\ {\rm yr}\ }{(\alpha+1)^{\frac{1}{6}}}\left(\frac{100\rm\ cm^{-3}}{n_0}\right)^{\frac{1}{3}}\nonumber\\
						&			&	\times\left(\frac{m_{\rm c}}{1{\rm M_{\oplus}}}\right)^{\frac{1}{3}}\left(\frac{r}{1.7\times10^{17}\rm\  cm}\right)^{\frac{2\alpha-1}{6}} .
		\end{eqnarray}
        
        		\noindent Both expressions were obtained under the assumption of hydrostatic equilibrium of the medium. 
        		Thus, the temperature profile follows $T(r)\propto r^{-1}$. 
        
\begin{figure*}
			\centering
			\includegraphics[width=0.495\textwidth]{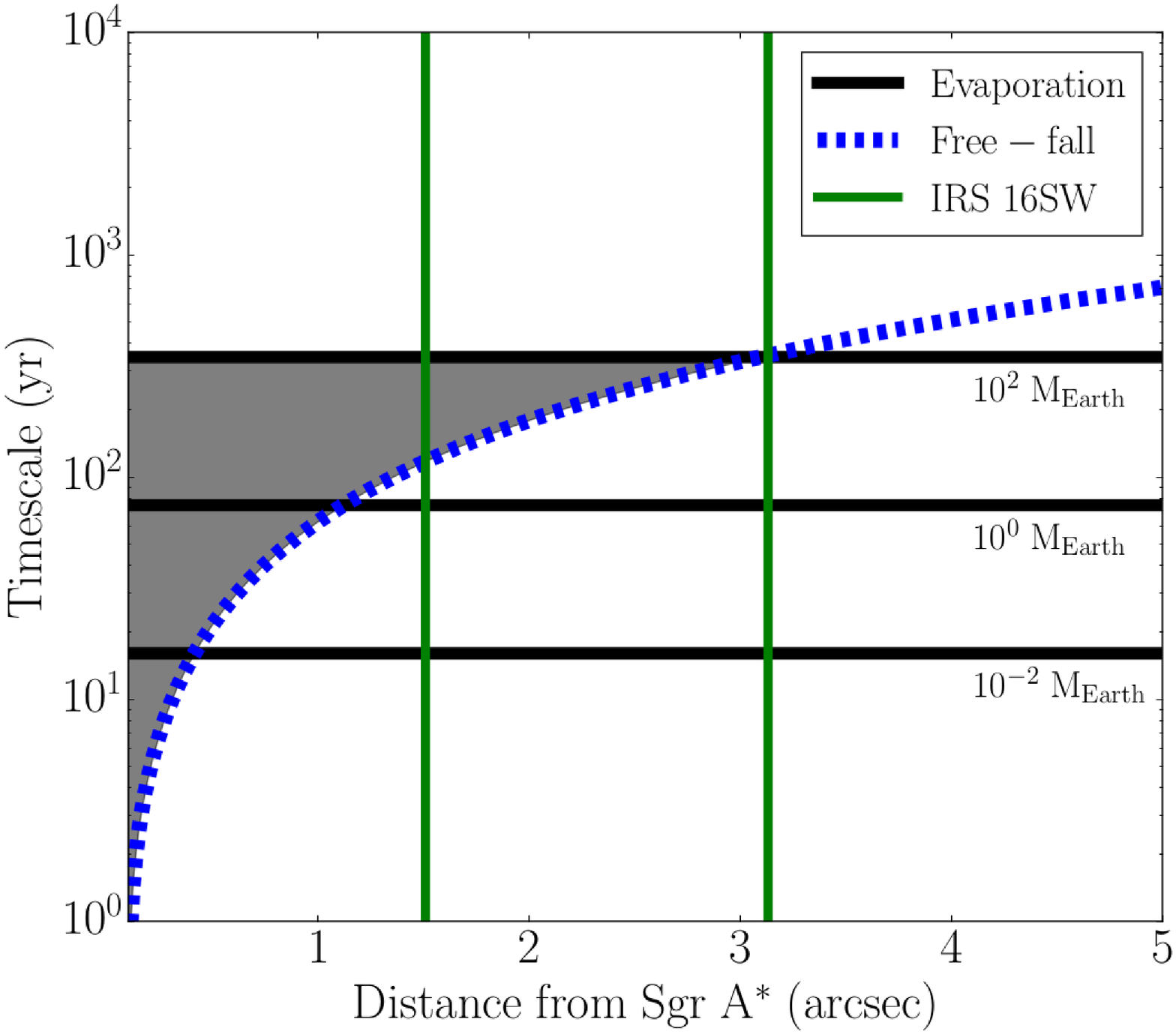}
			\includegraphics[width=0.495\textwidth]{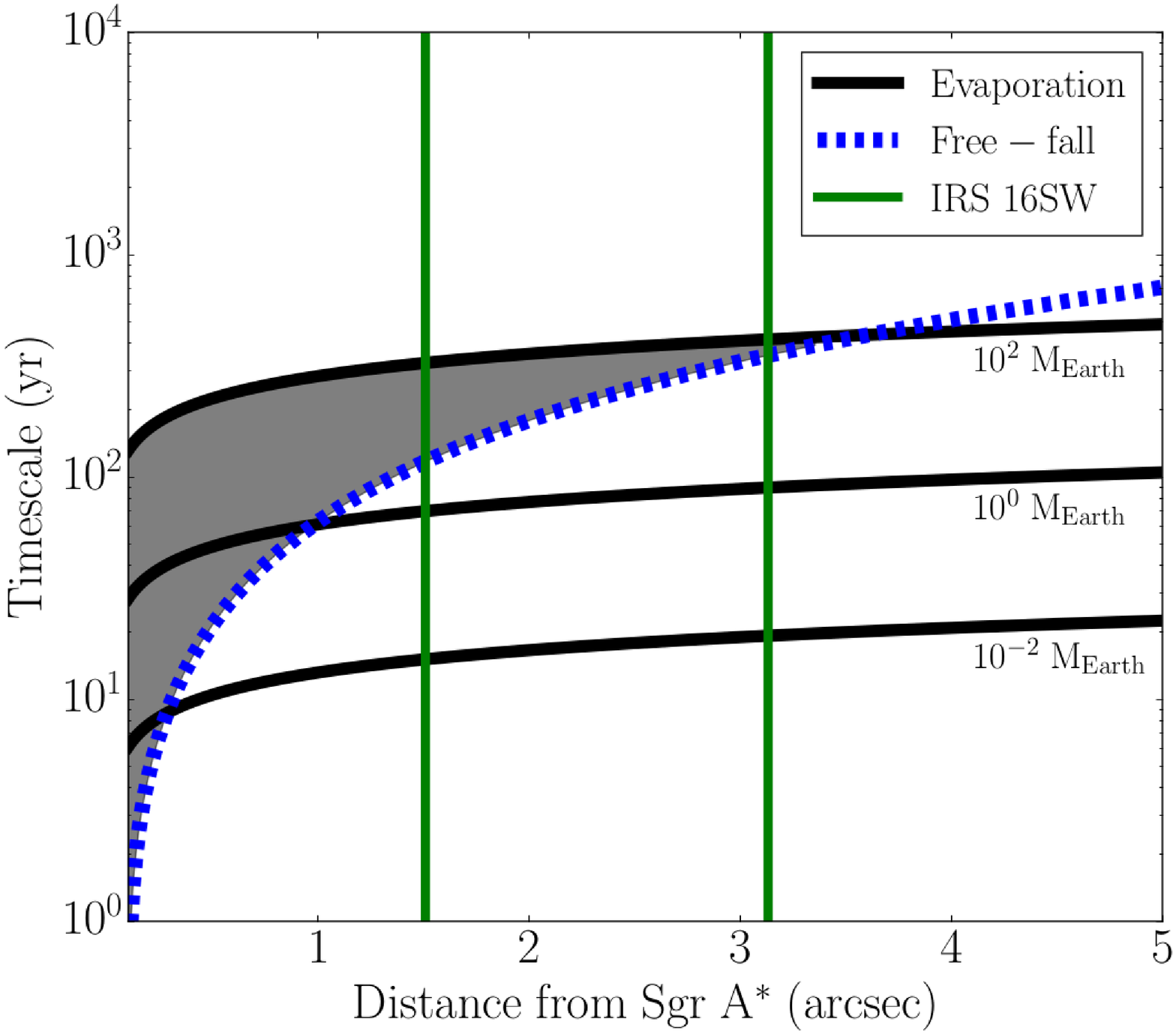}
			\caption{Relevant timescales as a function of distance from Sgr~A*. 
			The solid black lines show the evaporation timescale, i.e. lifetime, of gas clumps of a given mass embedded in a hot medium. 
			The dashed blue line represents the free-fall timescale. 
			The grey area shows the region where clumps of $M<100\rm\ M_{\oplus}$ need to be produced in order they could be captured. 
			The vertical solid green lines show the pericentre and apocentre distances of IRS~16SW orbit around Sgr~A*. 
			The left panel shows estimations using a radial density profile with $\alpha=0.5$, while the right panel uses $\alpha=1.5$.}
			\label{fig:timescales}
		 \end{figure*}

		In Figure~\ref{fig:timescales}, we show the evaporation timescale as a function of the distance from Sgr~A* for different clump masses (solid black lines). 
        		Also, we plotted the free-fall timescale (dashed blue lines), and the pericentre and apocentre distances of IRS~16SW (green vertical lines).
        		In the left panel, we present the case with $\alpha=0.5$ (shallow density profile) which describes an outflow solution for the accretion flow as suggested by \cite{W13}. 
        		In the right panel, we show the case with $\alpha=1.5$ (steep density profile) which represents a Bondi accretion flow environment. 
        		Despite having different dependences on $r$ the values of the evaporation timescale are very similar at the orbit of IRS~16SW. 
        		This is a direct consequence of both density models being scaled at the Bondi radius $r_{\rm b}\sim1.4\rm\ arcsec$, which corresponds to the typical distance between the binary and Sgr~A*. 
        		Thus at this separation both density profiles must converge.
        		On the contrary, approaching the SMBH models start to differ. 
        		The steep profile ($\alpha=1.5$) makes thermal conduction twice more efficient at $0.1\rm\ arcsec$ from Sgr~A* when comparing to the shallow profile ($\alpha=0.5$). 
        		This is the result of the medium becoming denser more rapidly as $r$ decreases in the case of the steep profile.  
        		Finally, notice that light clumps ($<1\rm\ M_{\oplus}$) will evaporate very rapidly but more massive ones ($>1\rm\ M_{\oplus}$) may survive for decades, and even centuries; regardless of the ISM profile. 
        
        		In Figure~\ref{fig:timescales}, we can observe it is not possible for light clumps ($<1\rm\ M_{\oplus}$) created by IRS~16SW to reach the SMBH indicated by the black solid lines being below the dashed blue line at distance of the orbit of IRS~16SW.
        		This means that the free-fall timescale is longer than the typical lifetime of the light clumps. 
        		Then, they will be evaporated very quickly without having chances of approaching to the SMBH.
        		Only fairly massive clumps ($\simgt3\rm\ M_{\oplus}$) live long enough to fall onto Sgr~A*. 
        		Therefore, only clumps roughly in the range $1-100\rm\ M_{\oplus}$ can achieve short distances to the central black hole. 
        		However, as we will discuss it is not likely for a binary like IRS~16SW to form such massive clumps (see Section~\ref{sec:mass}).
   
        		\cite{G12} estimated the gas mass of G2 from its Br$\gamma$ emission to be $\sim 3\rm\ M_{\oplus}$ assuming it is being kept at $10^4\rm\ K$. 
        		As clumps are evaporated constantly during their lives G2's initial mass had to be larger. 
        		Then, from the analysis of clump lifetimes it is still possible for G2 to be one of these gas clumps. 
        		However, this also puts strong constraints on the initial mass and the position where clumps have to be ejected from in order to mimic G2 observations. 

	\subsection{Clump masses}
	\label{sec:mass}
        
        		Here we will constrain the mass of clumps produced in a colliding wind binary. 
        		In principle, it is hard to predict the mass of clumps formed in a stellar wind collision. 
            	This is due to the fact that we need to derive their size to do so. 
            	In \cite{C16} we overcame this issue by using the unstable wavelength criteria of the NTSI as clump radius proxy. 
           	Nevertheless, such an approach can only be used when we can ignore wind acceleration, stellar gravity and radiation pressure. 
		In the case of IRS~16SW we cannot neglect those processes given the very short stellar separation of the binary of $\sim0.7\rm\ au$.
            	This is why here we follow a simpler but more appropriate approach instead.
            
            	Let us start by considering a system formed by two identical stars. 
            	Each of them blows a smooth, isotropic stellar wind whose density and terminal velocity are $\rho_{\rm w}(r)$ and $V_{\rm w}$, respectively (see Figure~\ref{fig:sketch1}). 
            	Then, the stellar wind density will be given by
            
            	\begin{equation}
            		\rho_{\rm w}(r)=\frac{1}{4\pi}\frac{\dot{M}_{\rm w}}{V_{\rm w}r^2},
            	\end{equation}
            
            	\noindent where $r$ is the distance to the star and $\dot{M}_{\rm w}$ its wind mass loss rate.
		Once winds collide, a slab of shocked material will be formed in the midpoint between the stars. 
            	Let us call the distance between a star and the slab $D$, i.e., the stellar separation is $2D$.
            	The slab will be confined by the ram pressure of the two winds. 
            	Assuming the gas within the slab cools down faster than it can escape from the system its density $\rho_{\rm s}$ can be estimated by
            
            	\begin{equation}
            		P_{\rm s}=P_{\rm w}\Rightarrow	\rho_{\rm s}=\frac{V_{\rm w}^2}{c_{\rm s}^2}\rho_{\rm w},
            	\end{equation}
            
            	\noindent where $P_{\rm s}$ is the thermal pressure within the slab, $P_{\rm w}$ is the ram pressure of the wind, and $c_{\rm s}$ is the slab sound speed.
            
            	As clumps are formed out of slab material their mass will be given by $m_{\rm c}\sim\rho_{\rm s} L^3$, where $L$ is the size of the clump. 
		Combining the previous expressions we can estimate the mass of the clumps as
            
            	\begin{equation}
            		m_{\rm c}\sim\frac{1}{4\pi}\frac{\dot{M}_{\rm w}V_{\rm w}D}{c_{\rm s}^2}\left(\frac{L}{D}\right)^3.
            	\end{equation}
            
		\noindent Notice that physically motivated clumps will satisfy $L/D<1$. 
		Otherwise, the clumps would overlap with the stars.
            	Therefore, we refer to clumps with $L>D$ as not physically motivated objects. 
            	Based on this, the maximum clump mass possible will be given when $L/D\sim 1$, i.e.,

             	\begin{equation}
			\label{eq:m_max}
            		m_{\rm c}< 3{\rm\ M_{\oplus}}\left(\frac{\dot{M}_{\rm w}}{10^{-5}\rm\ M_{\odot}\ yr^{-1}}\right) \left(\frac{V_{\rm w}}{600\rm\ km\ s^{-1}}\right)\left(\frac{D}{0.3\rm\ au}\right),
            	\end{equation}
            
            	\noindent where the sound speed within the slab was set to $10\rm\ km\ s^{-1}$. 
            	This corresponds to the sound speed of an ionised ideal gas at $10^4\rm\ K$. 
            	This value is the temperature floor set by the UV radiation of the massive stars.
            	The other quantities were scaled by the values of the binary IRS~16SW \citep{M06,C08}. 
            	It is important to remark that our assumptions are valid for this binary: identical stars with radiative winds \citep{M06,C16}.
            
            	Thus, this result shows that massive clumps cannot be produced in IRS~16SW. 
            	In reality, it is possible that this limit is even smaller.  
            	The short stellar separation of the binary (about $0.7\rm\ au$) is such that it is probable stellar winds do not reach terminal velocity before they collide. 
            	As a consequence, the actual upper limit will be only a fraction of our estimate, making the formation of a clump as massive as G2 less likely.
            
            	In principle, this result rules out the possibility of a G2-like clump to be produced in IRS~16SW. 
            	The fact that clumps are constantly losing mass through thermal conduction will reduce their initial mass since the moment they are born. 
            	Currently, G2's observed gas mass is very similar to the maximum clump mass produced in IRS~16SW. 
            	Therefore, for G2 to have been born in the binary it should not have lost mass through its life. 
            	This statement is very unlikely due to the large temperature difference between G2 and the ISM. 
            	However, the process of clump formation in colliding wind binaries is not well studied yet. 
            	It still remains as an option that a collection of smaller clumps were ejected and traveled together to Sgr~A*. 
            	If this is the case they could be observed as a single larger clump (see \S\ref{sec:rate}). 
	
		In the case of the other binaries in the region, we can use Equation~\ref{eq:m_max} to calculate the most massive clump they could create. 
		From this, we obtain that IRS~16NE could generate at most clumps with a mass of $30\rm\ M_{\oplus}$. 
		Although this is ten times larger than IRS~16SW clumps, the fact that the system is roughly twice as far from Sgr~A* makes it impossible for clumps to survive until reaching the SMBH (see Figure~\ref{fig:timescales}). 
		The scenario is not favourable to E60 either. 
		Its clumps are expected to be at most $0.7\rm\ M_{\oplus}$. 
		As this binary is further away than IRS~16SW, clumps cannot survive for long enough to fall into the very centre. 
		Thus, clumps produced in these systems are less likely to reproduce the G2 source.
        
        \subsection{The clump ejection rate of IRS 16SW}
	\label{sec:rate}
	
\begin{figure}
			\centering
			\includegraphics[width=0.45\textwidth]{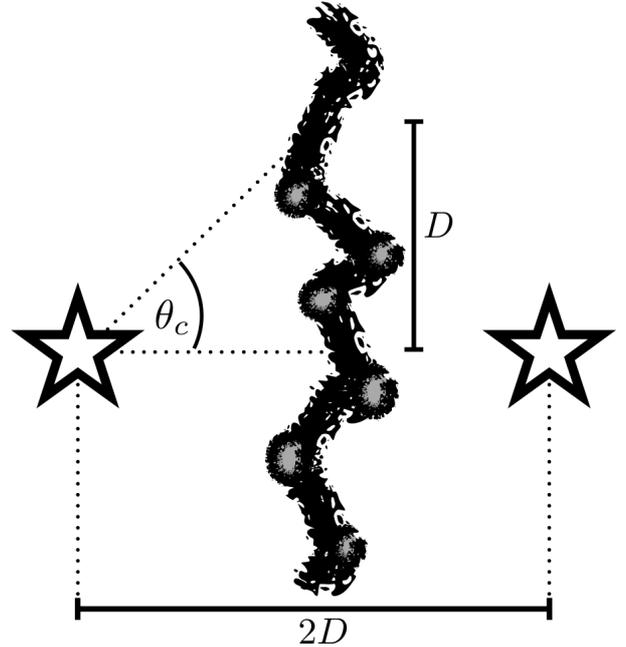}
			\caption{
			Schematic representation of an unstable colliding wind system. 
			A gas slab is formed after the wind collision which is illustrated as the vertical thick sinusoidal line.
            		If it becomes unstable clumps will be created. 
			Clumps are represented as grey knots located in the slab.
			Overdensities formed as a result of the NTSI are expected to be formed close to the line connecting the two stars. 
			The vertical length $D$ is shown to illustrate the region where we consider clumps are created.}
			\label{fig:sketch1}
		\end{figure}

		In this section we proceed to estimate the amount of clumps a colliding wind binary can create given its properties. 
		We start assuming clumps are exclusively created through the NTSI. 
        		This assumption is justified as numerical simulations have shown this instability dominates over others on long timescales \citep{L11}.
		In this context, clumps are formed very close to the line connecting both stars. 
		Therefore, calculating the amount of material flowing into this region we can constrain the mass that could be transformed into clumps per unit of time. 
		In order to delimitate this area we will make use of the length scale of the system, i.e., the stellar separation. 
		So, we will assume clumps are formed in a portion of the slab of length $2D$ as shown in Figure~\ref{fig:sketch1}. 
		Then, we proceed to estimate the mass flux that goes into a solid angle subtended by $D$ observed from the position of one of the stars (and assuming azimuthal symmetry). 
		Let us define $\theta_c$ as the polar angle as shown in Figure~\ref{fig:sketch1}. 
		Its value can be easily calculated if we assume both stellar winds are identical.  
		Therefore, the slab will be located exactly in the midpoint between the stars.  
		Thus, $\theta_c=\arctan(1)=\frac{\pi}{4}$.
		Now, we take the ratio of the solid angle covered by a patch on sky of one of the stars described by $0\leq\theta\leq\theta_c$ and $0\leq\phi\leq2\pi$, and the whole sky. 
		Then, 
	
		\begin{equation}
			\frac{S_c}{S_{\rm sky}}=\frac{1}{4\pi}\int^{2\pi}_0\int^{\theta_c}_0\sin(\theta)d\theta d\phi\approx15\%. 
		\end{equation}
		
		\noindent This means that clumps would be formed within 15\% of the sky of each star. 
		From spectral analysis we know that the mass loss rate of IRS~16SW is $10^{-5}\rm\ M_{\odot}\ yr^{-1}$ \citep{C08}. 
        		Thus, about $10^{-6}\rm\ M_{\odot}\ yr^{-1}$ will be transformed into clumps according to our estimate. 
		Then, the rate at which clumps will be formed is $\dot{M}_{\rm clump}\sim0.4\rm\ M_{\oplus}\ yr^{-1}$. 
        		This means it would take about eight years to create 3-$\rm M_{\oplus}$ in clumps. 
        		If the whole mass can go into a single clump the binary would need $\sim140$ binary periods to create such a clump. 
        		This is not realistic because it is significantly longer than the dynamical timescale of the system. 
        		Even in the extreme case in which the complete mass loss rate goes into clumps it would take about 14 binary periods to form a single G2-mass clump. 
        		Thus, it is very hard to reconcile the idea of creating a clump with G2's mass.
		
\section{Clump ejection along IRS 16SW orbit and simulation setup}
	\label{sec:ejec}
    
        Up to now we have investigated properties of clumps formed in wind collisions. 
        However, to be able to test more quantitatively whether G2 could be one of such clumps we need to compare its observations (positions and velocities) with the output of our model.
        Specifically, here we study if G2's dynamics can be reproduced with a clump ejected from IRS~16SW. 

    	Our approach consists in describing the motion of clumps ejected from IRS 16SW while orbiting the SMBH.
   	Specifically, we compute their trajectories in the presence of the gravitational field of Sgr~A*, and the drag force exerted by the ISM.
    	Furthermore, we considered clump initial velocities and their finite lifetimes given by thermal conduction effects previously discussed.  
    	Firstly,  we describe our model and describe the impact of different input parameters. 
    	In the next Section, we present the results of this analysis.  
	
	\begin{figure*}
		\centering
		\includegraphics[width=0.9\textwidth]{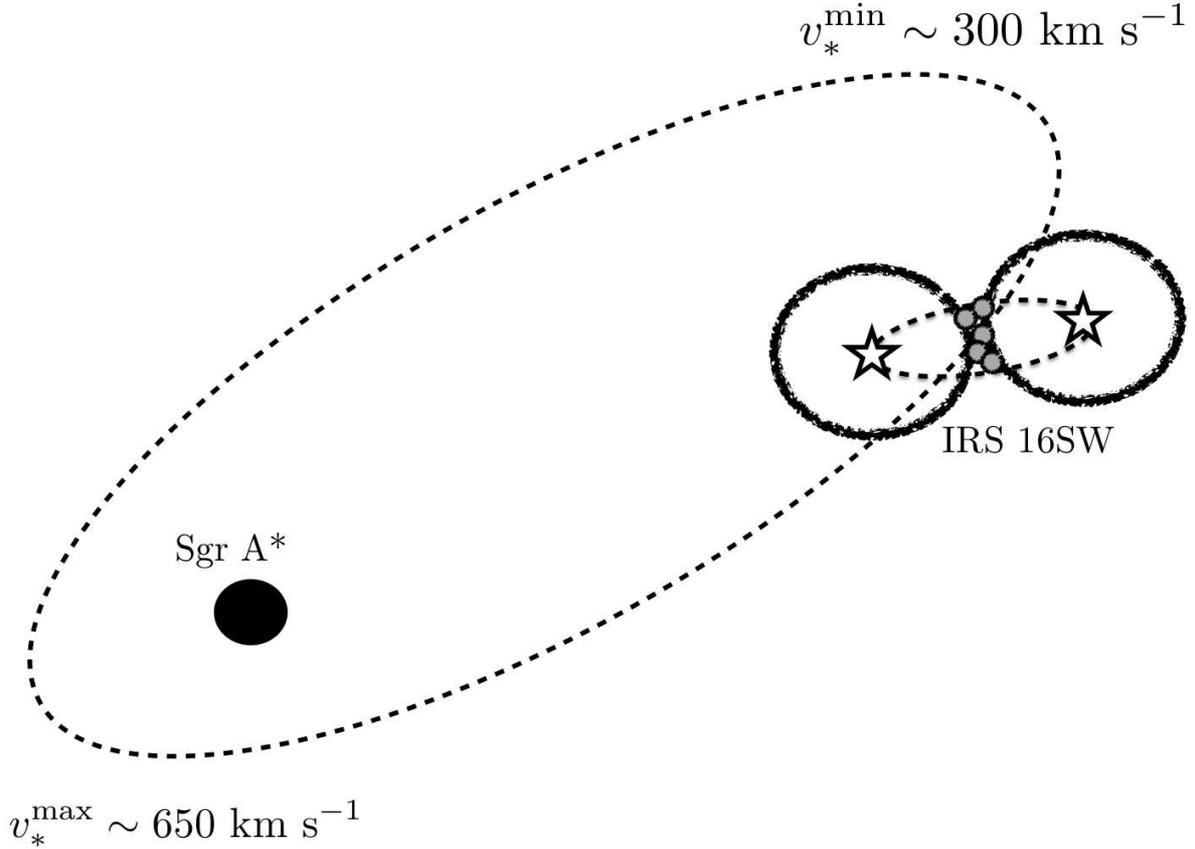}
		\caption{Schematic representation of the binary system IRS~16SW orbiting Sgr~A*. 
		The stellar components of the binary are expected to blow strong winds. 
		In the collision region of such outflows clump formation takes place.}
		\label{fig:sketch2}
	\end{figure*}
	
	\subsection{Clump equation of motion and simulation setup}
	
		We start considering IRS~16SW ejects clumps isotropically while it is orbiting around Sgr~A* (see Figure~\ref{fig:sketch2}). 
		Every single clump ejected will have an initial velocity with a random orientation.
		We discuss the impact of this assumption in Section~\ref{sec:lim}.
		Immediately after being ejected clumps are subject to the gravitational pull of the SMBH. 
		In principle, this causes clumps to move on Keplerian orbits. 
		However, we included the presence of drag exerted by the ISM that will cause deviations from such Keplerian motion. 
		We modelled this interaction as ram pressure considering the relative motion of the medium and the gas clump. 
		Then, the equation of motion of each clump will be given by
	
		\begin{equation}
			\label{eq:drag}
			\frac{d^2\vec{r}}{dt^2}=-\frac{GM_{\rm BH}}{r^2}\hat{r}-\frac{\sigma_{\rm c}(t)}{m_{\rm c}(t)}\rho_{\rm ISM}(r)\left|\frac{d\vec{r}}{dt}\right|^2\hat{v},
		\end{equation}
	
		\noindent where $G$ is the gravitational constant, $M_{\rm BH}$ is the mass of Sgr~A*, $\rho_{\rm ISM}$ is the density of the ISM,  $\hat{r}$ is the unit vector in the radial direction, $\hat{v}$ is the unit vector in the velocity direction, $m_{\rm c}$ and $\sigma_{\rm c}$ are the clump mass and cross section, respectively. 
        
		The initial conditions will be given by
	
		\begin{eqnarray}
			\frac{d\vec{r}}{dt}(t=0)		&	=	&	\vec{v}_* + \vec{v}_{\rm ej}\\
			\vec{r}(t=0)			&	=	&	\vec{r}_*
		\end{eqnarray}
		
		\noindent where $\vec{v}_{\rm ej}$ is the initial velocity kick whose direction is random, $\vec{r}_*$ and $\vec{v}_*$ are the binary system position and velocity vectors, respectively.
        
       		Furthermore, we included the clump mass loss due to thermal conduction which is the main constraint on clump lifetimes. 
       		This can be expressed by 
        
        		\begin{equation}
        			\frac{dm_{\rm c}}{dt}=-\dot{M}_{\rm evap}(m_{\rm c},r),
        		\end{equation}
        
        		\noindent where $\dot{M}_{\rm evap}(m_{\rm c},r)$ is given in Equation~\ref{eq:mdot}
        
		The model has four input parameters: initial clump mass $m_{\rm c}(t=0)$, ejection rate (number of clumps ejected per time unit), the clump initial kick velocity $\vec{v}_{\rm ej}$, and the density profile power law $\alpha$. 
		Once we set those parameters we compute the trajectory of every single clump solving Equation~\ref{eq:drag}.
        		It is important to remark that in our model clump ejection occurs while IRS~16SW orbits around Sgr~A*.
        		This means clumps are ejected from different locations on the binary orbit.
             
        		Our simulations are run for a period of $300\rm\ yr$ which is longer than the lifetime of any clump IRS~16SW can create (see Section~\ref{sec:mass}). 
        		Therefore, simulations start and finish in $t=1716\rm\ yr$ and $t=2016\rm\ yr$, respectively. 
		Finally, from every run we registered clumps whose mass is at least 10\% of Earth's at the present, i.e., $m_{\rm c}(t=2016\rm\ {\rm\ yr})\geq0.1\ M_{\oplus}$.  
		Then, we can get an estimate of the spatial distribution of gas clumps present in the region. 
        
        		In order to avoid undesired statistical noise we consider a very large clump ejection rate.
        		Then, we normalise the number of counts output over the total number of clumps ejected.
		Therefore, the outputs of the model are \textit{clump fractions} $f$, i.e., number of clumps whose $m_{\rm c}>0.1\rm\ M_{\oplus}$ divided by the total number of ejected clumps. 
		However, we are mostly interested only in clumps that could reproduce G2's motion. 
        		Thus, we define the \textit{G2 candidate fraction} $f_{\rm G2}$ as the number of clumps that satisfy $m_{\rm c}>0.1\rm\ M_{\oplus}$ and $r(t=2016\rm\ yr)\leq 0.5\rm\ arcsec$ divided by the total number of clumps ejected.  
        		Then, to obtain an absolute value out of this we used the mass in clumps created calculated in Section~\ref{sec:rate}. 
		
		Before jumping to the results, let us bear in mind that the model had three input parameters: clump mass $m_{\rm c}$, ejection speed ${v}_{\rm ej}$,  and ISM density profile power-law $\alpha$. 
    		The choice of a very large value of the clump ejection rate allowed us to reduce the number of parameters. 
    		Furthermore, instead of using a single clump mass value, we considered a clump mass function as a semi-log distribution. 
    		Here, the mean and standard deviation need to be specified. 
    		Although we modelled different mean values, the standard deviation was fixed to $0.1\rm\ dex$ in all simulations.
    		This choice is inspired by our forthcoming work based on numerical hydrodynamical simulations of clump formation in stellar wind collisions (Calder\'on et al.; in preparation). 
	
    		The trajectory of IRS~16SW between years 1716 and 2016 was sampled using one thousand points evenly spaced in time. 
    		On each of those points 2048 clumps are ejected isotropically with a speed $v_{\rm ej}$ (in the reference frame of the binary). 
		The ejection speed values used are fractions of the terminal velocity of the stellar wind of IRS~16SW. 
		All these values are within the range of the orbital speed of the binary around Sgr~A*, i.e., $300-650\rm\ km\ s^{-1}$.  
    		The orbital sampling and clump ejection number were selected in order that the statistical noise of the G2 candidate fraction $f_{\rm G2}$ was kept to less than $5\%$. 
		In the following section we present the G2 candidate fraction of each set of simulations, the trajectories of such clumps, the role of the drag force on such trajectories, and the expected clump mass distribution.

	\subsection{Simulation results}
	\label{sec:res}
        
		\subsubsection{G2 candidate fraction}

    			In Table~\ref{tab:runs}, we present the parameters used in our simulations, as well as diagnostics of their results.  
    			Runs with clump mass functions whose mean mass was smaller than a tenth of an Earth-mass did not register any G2 candidate, i.e., $f_{\rm G2}=0$,  as expected from our analytical estimates. 
    			Different ejection speeds or density profiles did not change this result. 
    			Basically, those clumps have lifetimes that are too short for being able to travel close to Sgr~A*.
    			Only clumps with initial masses higher than a single Earth-mass have chances of being captured regardless of the ejection speed and density profile (see Figure~\ref{fig:timescales}).
    
    			In the next set of simulations we used a mean clump mass of $\bar{m}_{\rm c}=3\rm\ M_{\oplus}$. 
   			This is the highest physical mass a clump produced in IRS~16SW can have (see Section~\ref{sec:mass}). 
    			As we expect more massive clumps to live for longer, this choice should maximise the G2 candidate fraction. 
    			The results are also shown in Table~\ref{tab:runs}. 
   			Here, we observe that G2 candidate fractions are in all cases in the range $1.0 - 1.3\%$, with a shallow maximum for ejection speeds $350-400\,$km$\,$s$^{-1}$. 
    			Higher ejection speeds make clumps more likely to get on unbound orbits. 
    			On the contrary, a slower initial speed would keep more clumps bound to the SMBH. 
			However, decreasing the initial speed also causes clumps to retain more angular momentum making infall less likely to occur. 
			Although the drag could eventually place them on more radial orbits, their limited lifetimes do not allow this to happen very easily.
    			We see no difference in the results when using different density profiles for the medium. 
    			In principle, this is expected due to the similarity between both profiles at the distance of the orbit of IRS~16SW, specially on the effects of thermal conduction (see Figure~\ref{fig:timescales}).
    			If we consider a clump ejection rate of $0.4\rm\ M_{\oplus}\ yr^{-1}$ (see Section~\ref{sec:rate}) the ejection rate of 3-$M_{\oplus}$ clumps will be $0.13\rm\ yr^{-1}$. 
    			Thus, in $300$ years $\sim40$ clumps should be ejected. 
    			As $\approx 1\%$ of them are G2 candidates, we should expect none, or at most one. 
    			Could G2 then correspond to a clump created in this way?  
			Our simulations can give us more information, from a dynamical point of view. 
    			In the next section we will check if the trajectory of such clumps can mimic the observed motion of G2.
    
    			\begin{table*}
			\begin{threeparttable}
				\centering
        				\caption{Input parameters and results of simulation runs. 
        				In all simulations $\sim2\times10^6$ clumps were ejected per orbit.  
        				Also, the clump mass function standard deviation was fixed to 0.1~dex.
        				Notice that no clumps were captured in the low mass clump runs.}
    				\begin{tabular}{|c|c|c|c|c|c|}
        		 			\hline
        		 			Name	&	$\bar{m}_{\rm c}$ $(M_{\oplus})$		&	$|v_{\rm ej}|$ $(\rm\ km\ s^{-1})$		&	$\alpha$	&	$f_{\rm G2}$	&	$\min(\chi^2_{\rm d.o.f.})$\\
        			 		(1)				&	(2)		&	(3)		&	(4)		&	(5)		&	(6)\\
        		 			\hline
        		 			\hline
        		 			Low mass			&	$<0.1$	&	any		&	any		&	0		&	--\\
        	 				M1				&	1		&	any		&	any		&	$<0.1\%$	&	--\\
         				\hline
         				M\_G2\_v300\_a05	&	3		&	300		&	0.5	&	$1.2\%$	&	$2351$\\
         				M\_G2\_v300\_a15	&	3		&	300		&	1.5	&	$1.1\%$	&	$2101$\\
         				\hline
         				M\_G2\_v350\_a05	&	3		&	350		&	0.5	&	$1.3\%$	&	$815$\\
         				M\_G2\_v350\_a15	&	3		&	350		&	1.5	&	$1.2\%$	&	$715$\\
         				\hline
         				M\_G2\_v400\_a05	&	3		&	400		&	0.5	&	$1.3\%$	&	253\\
         				M\_G2\_v400\_a15	&	3		&	400		&	1.5	&	$1.2\%$	&	340\\
         				\hline
         				M\_G2\_v450\_a05	&	3		&	450		&	0.5	&	$1.2\%$	&	411\\
         				M\_G2\_v450\_a15	&	3		&	450		&	1.5	&	$1.2\%$	&	376\\
         				\hline
         				M\_G2\_v500\_a05	&	3		&	500		&	0.5	&	$1.1\%$	&	633\\
         				M\_G2\_v500\_a15	&	3		&	500		&	1.5	&	$1.0\%$	&	566\\
         				\hline
				\end{tabular}
				\label{tab:runs}
				\begin{tablenotes}
					\item \textit{Notes.} 
					Column 1: model ID. 
					Column 2: mean mass of the initial clump mass function. 
					Column 3: clump ejection speed (in the reference frame of the binary). 
					Column 4: power-law of the density profile of the medium. 
					Column 5: G2 candidate fraction, i.e., fraction of clumps whose mass satisfy $m_{\rm c}(t=2016\rm\ yr) >0.1\rm\ M_{\oplus}$ and are located within half arcsecond from Sgr~A*.
					Column 6: reduced chi-square value of the clump that is the best-fit to G2 observations. 
				\end{tablenotes}
			\end{threeparttable}
			\end{table*}

		\subsubsection{Captured clump trajectories}
    		
    			In our simulations we computed the trajectories of clumps up to the epoch of the latest G2 observations available in the literature \citep{P17}. 
        			It is important to remark that we excluded clumps whose positions at $t=2016\rm\ yr$ are further than half arcsecond from Sgr~A*. 
        			Then, we registered state vectors of clumps, i.e., position and velocity vectors at the same time at which observations took place. 
        			With this information we computed a Keplerian orbit for each clump. 
        			From this we compared the observed positions on the sky and line-of-sight velocities at the same epochs of the observations of G2\footnote{
			Observational data points were obtained sampling the posterior distribution of the model of \cite{P17} which reproduces extremely well the G2 dynamics.  
			We also made the analysis using the previously published data points \citep{P15} plus the latest epochs taken from the posteriors and found no significant differences.}.
        			Thus, we estimated the $\chi^2_{\rm d.o.f.}$ for each captured clump.

\begin{figure*}
				\centering
				\includegraphics[width=0.49\textwidth]{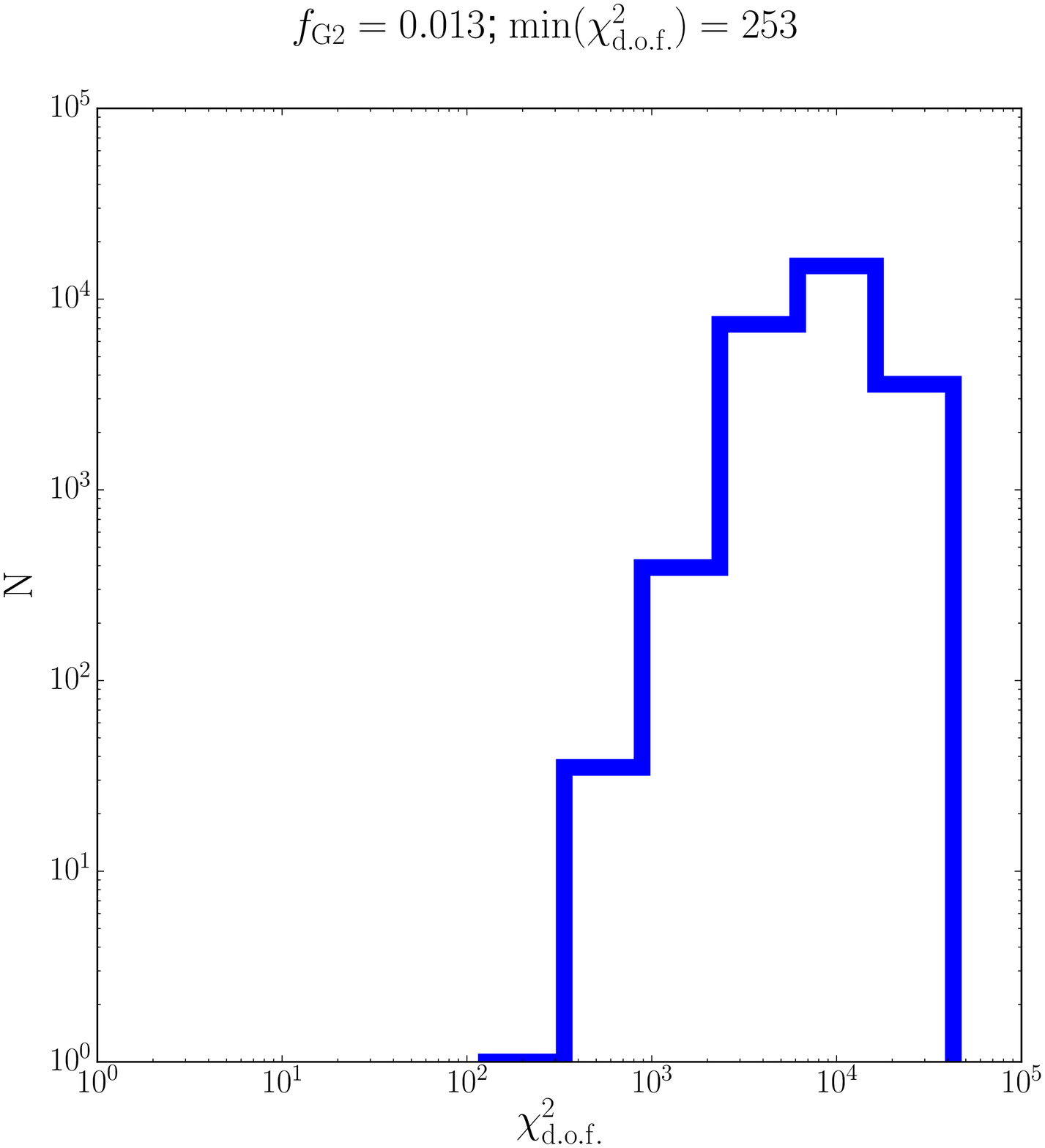}
           		 	\includegraphics[width=0.49\textwidth]{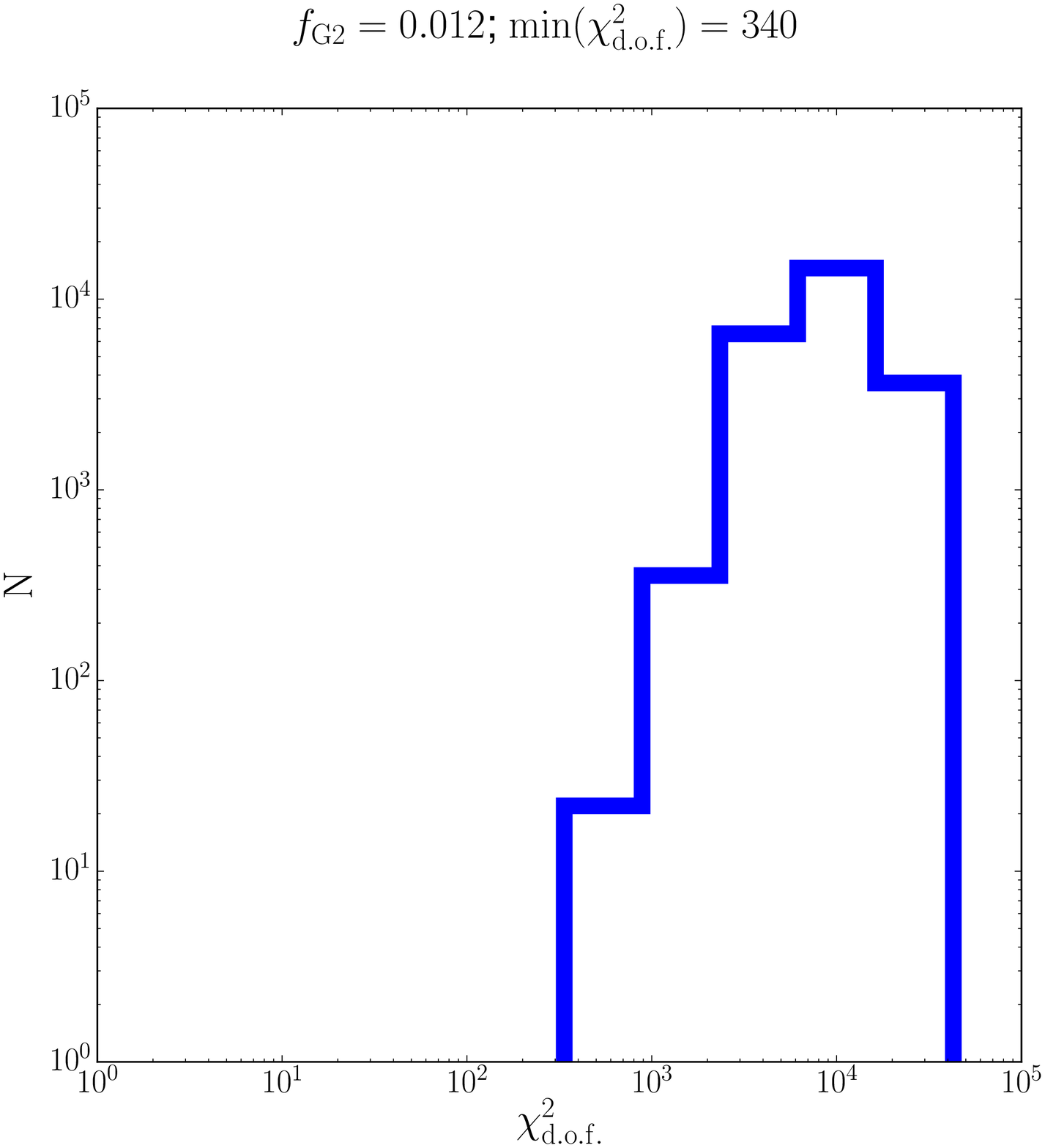}
				\caption{Histograms of the $\chi^2_{\rm d.o.f.}$ calculated for each G2 candidate using sky-positions and the line-of-sight velocities. 
				Left and right panels show the results of models M\_G2\_v400\_a05 and M\_G2\_v400\_a15, respectively.  
				The minimum $\chi^2_{\rm d.o.f.}$ value is shown on top of each plot along with $f_{\rm G2}$.}
				\label{fig:chi2}
			\end{figure*}

       			Figure~\ref{fig:chi2} shows histograms of the $\chi^2_{\rm d.o.f.}$ distribution for two of our simulations: M\_G2\_a05\_v400 and M\_G2\_a15\_v400. 
        			They correspond to models with  ($\alpha=0.5$, $v_{\rm ej}=400\rm\ km\ s^{-1}$), and ($\alpha=1.5$, $v_{\rm ej}=400\rm\ km\ s^{-1}$), respectively. 
        			We do not show histograms of other runs because in those cases the minimum $\chi^2_{\rm d.o.f.}$ was even larger (see Table~\ref{tab:runs}).
        			Notice that in both cases the $\chi^2_{\rm d.o.f.}$ values are significantly higher than unity. 
        			Thus, all G2 candidates are far from reproducing G2 observed properties.

\begin{figure*}
				\centering
				\includegraphics[width=0.495\textwidth]{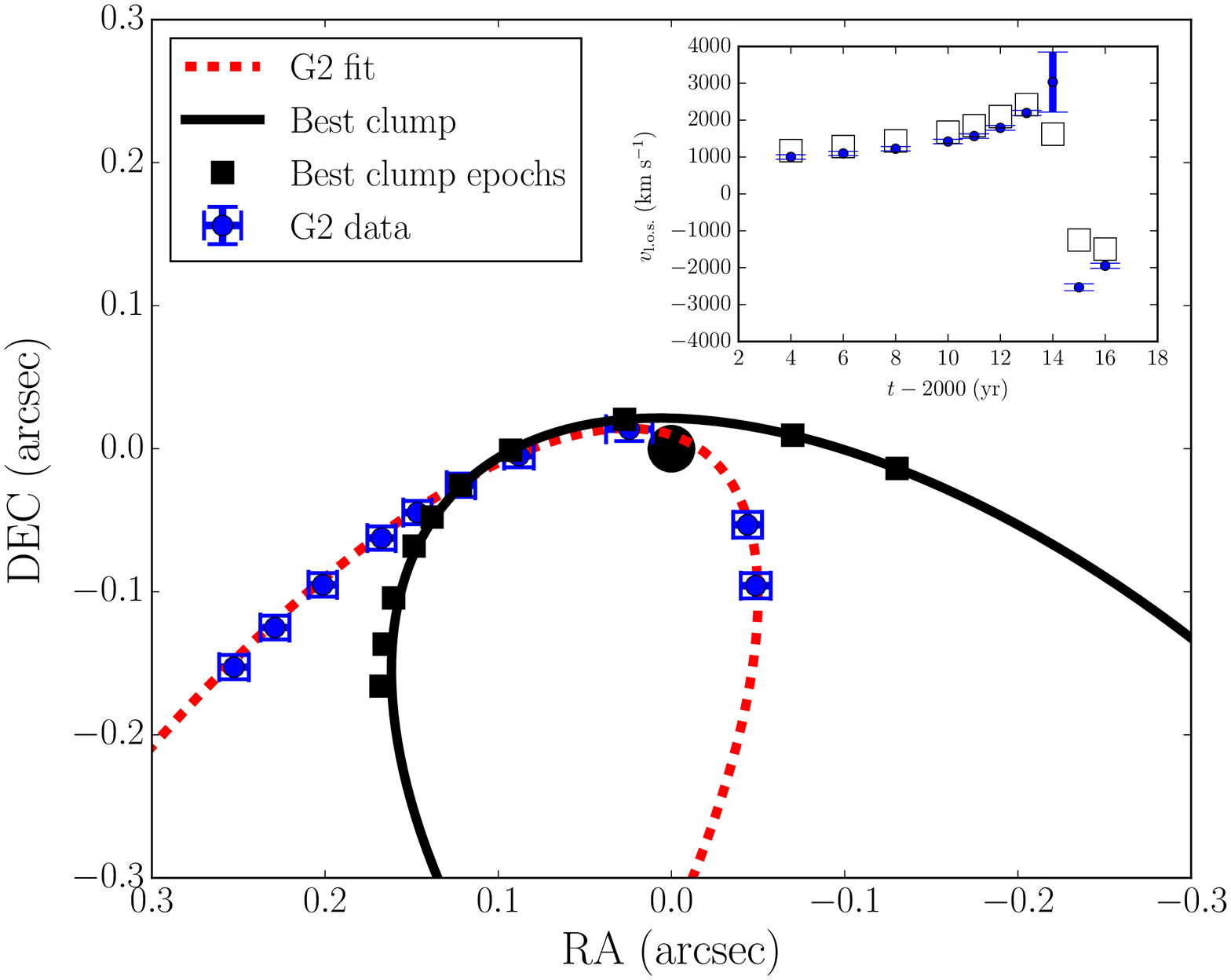}
            			\includegraphics[width=0.495\textwidth]{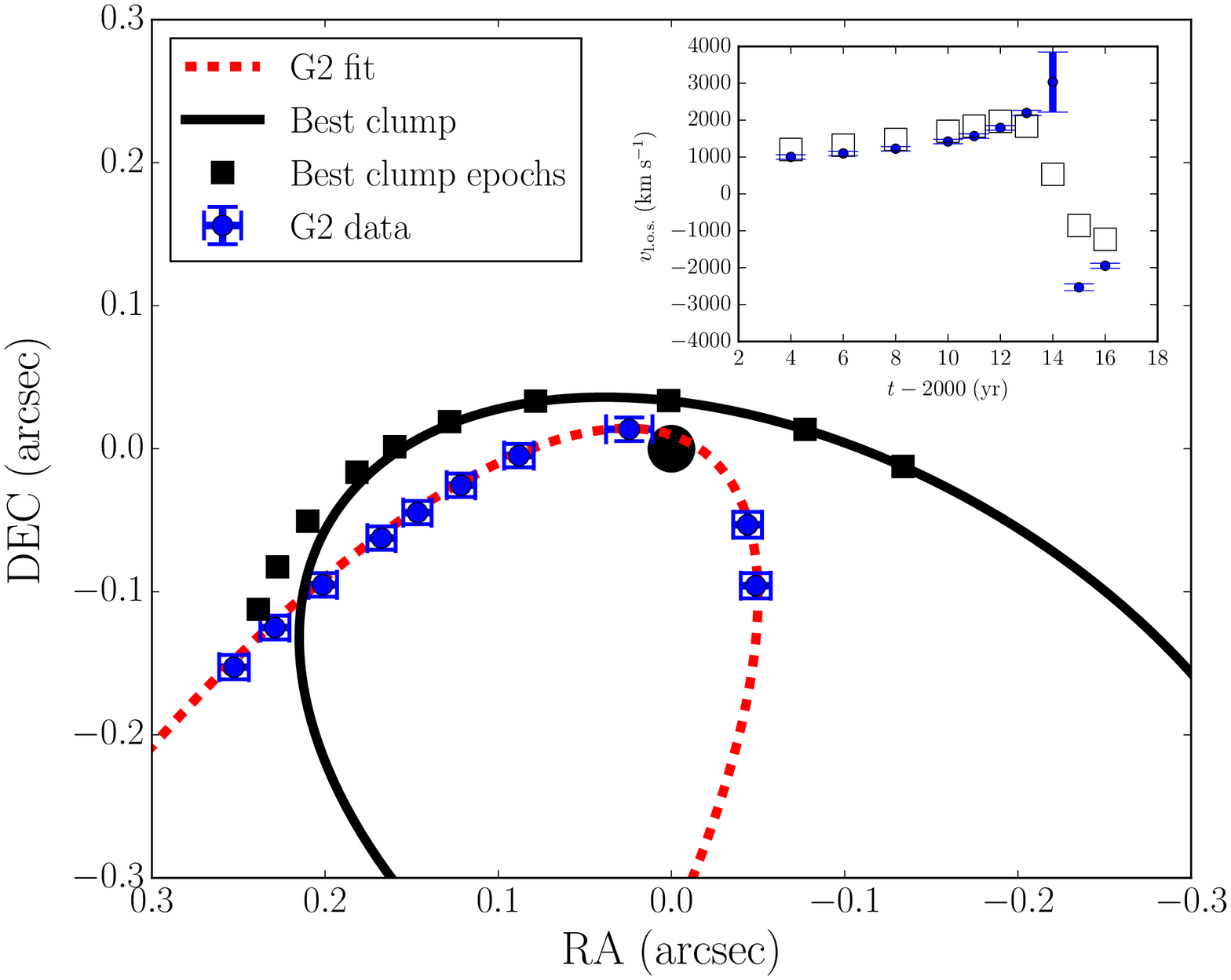}
				\caption{Sky projected positions and orbital fits for G2 and the clump whose $\chi^2_{\rm d.o.f.}$ is the smallest in a given model. 
				Left and right panels show the results of models M\_G2\_v400\_a05 and M\_G2\_v400\_a15, respectively. 
				Blue circles with error bars are the observed positions of G2. 
				Black rectangles show the positions of the best-fit clump at the same epochs as the G2 observations. 
				The dashed red and solid black lines stand for the Keplerian orbital fits to G2 data and to the last epoch of the modelled clump, respectively. 
				The inset graphs show the line-of-sight velocity as a function of time, where symbols retain their meaning, i.e., blue circles and black squares stand for G2 observations and best clump model, respectively. 
				}
				\label{fig:best_g2}
			\end{figure*}

	        		In each panel of Figure~\ref{fig:best_g2}, we show the orbit of G2 (dashed red line) and the clump whose $\chi^2_{\rm d.o.f.}$ is the minimum (solid black line) of models M\_G2\_a05\_v400 and M\_G2\_a15\_v400.  
        			Furthermore, the figure includes sky position measurements of G2 with their respective error bars (blue circles). 
       		 	As comparison we show clump positions on the sky at the epochs of the observations (black squares). 
        			The inset of each panel shows the line-of-sight velocity measurements of G2 (blue circles) and the clump (black squares). 
			We also show the clump Keplerian orbit computed from its most recent state vector (solid black line). 
			This allows us to visualise deviations of the clump orbits from Keplerian motion, since at earlier epochs the clump position does not follow exactly the Keplerian orbit.

        			From the analysis of Figure~\ref{fig:best_g2}, it is clear that G2 observations differ significantly from our models. 
        			Although some points fall within the error bars of the observations most of them display significant deviations from the data. 
        			The main differences seem to be on the orientation where the ellipse is pointing. 
        			Thus, observations already constrain strongly the direction from where G2 seems to be coming from. 
        			None of IRS~16SW clumps can mimic this constraint. 
        			This issue was discussed previously in Section~\ref{sec:kepler}. 
        			However, in the present analysis we did not specify any orbital fit to G2. 
        			Instead we compared a set of data points, and we ended up reaching the same conclusion. 
        			Thus, the observed motion of IRS~16SW and G2 are not consistent with the cloud originating from the binary. 
        
		\subsubsection{Role of the drag force}
     		\label{sec:fdrag}
     
     			In order to understand the impact of the drag force on clump trajectories we ran an extra set of simulations without including it. 
        			We made the same analysis of the outputs, i.e., register G2 candidates and check if their trajectories fit observations. 
        			The results do not show differences on the fraction of G2 candidates $f_{\rm G2}$. 
        			Although some differences were obtained when comparing clump and G2 motions they are not relevant as the $\chi^2_{\rm d.o.f.}$ did not change significantly. 
        
\begin{figure*}
				\centering
				\includegraphics[width=0.495\textwidth]{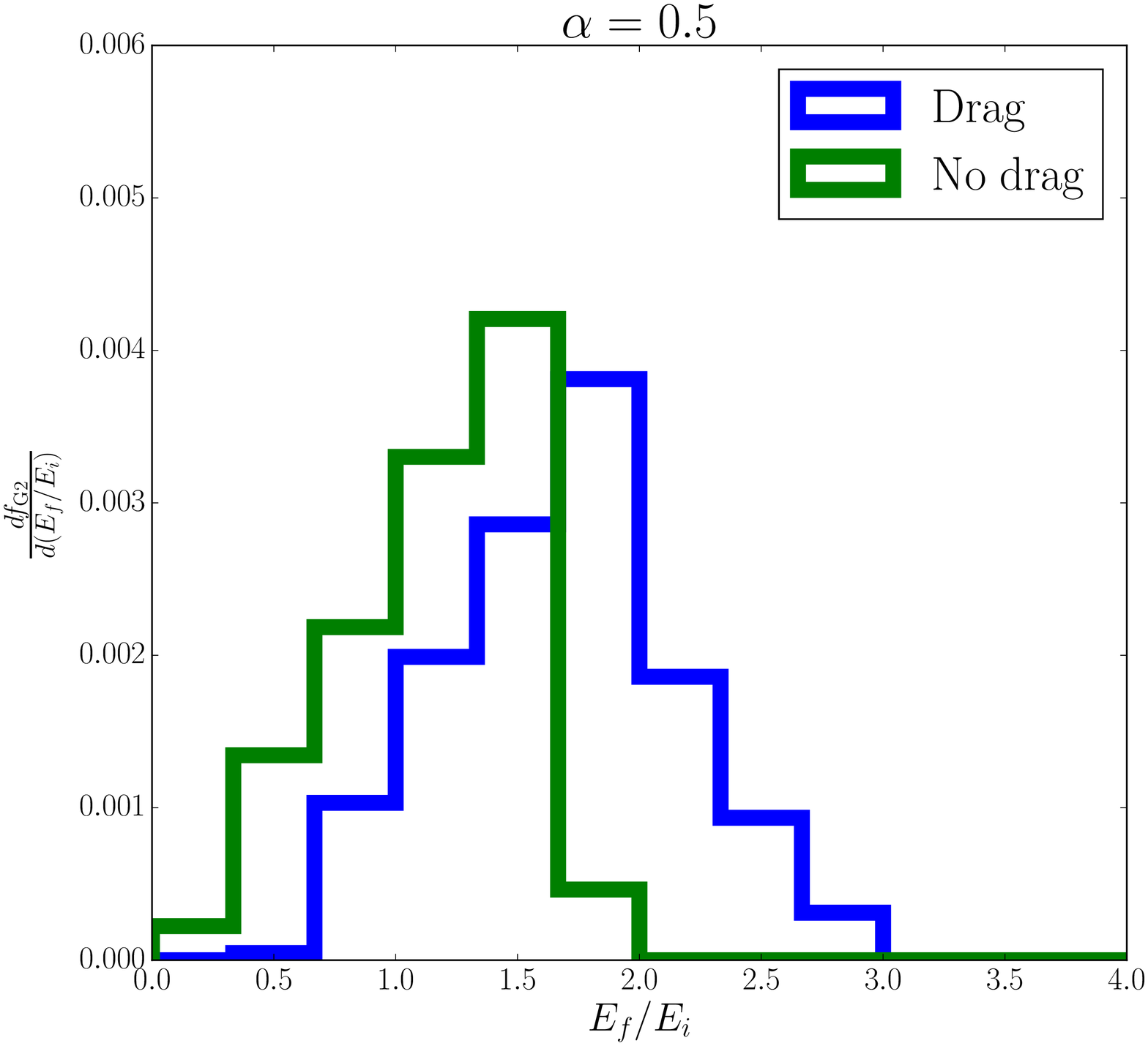}
            			\includegraphics[width=0.495\textwidth]{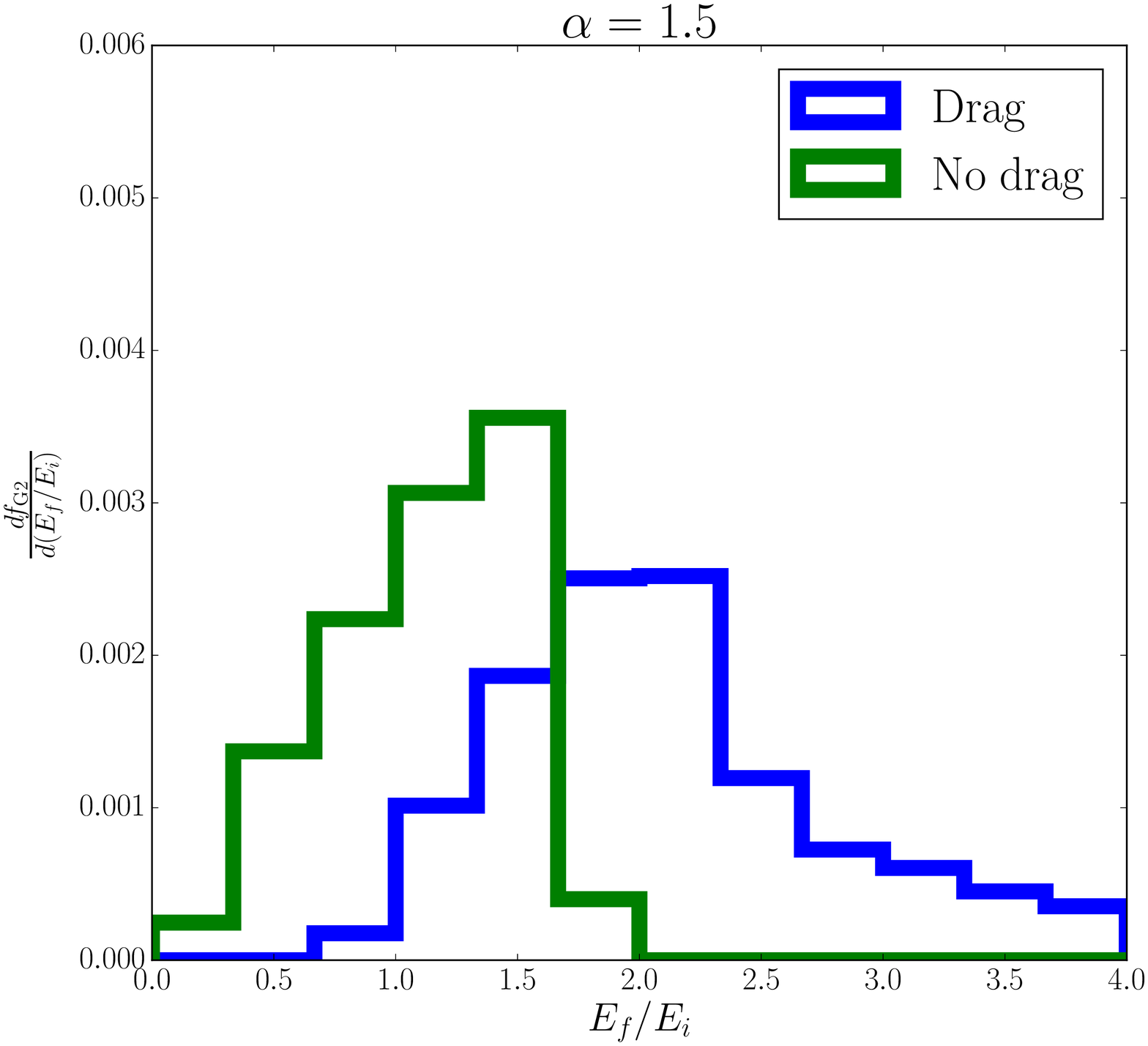}
				\caption{Histograms of the final binding energy of G2 candidates scaled by the orbital energy of IRS~16SW, i.e., $E_i$. 
            			Counts were normalised by the total number of clumps ejected. 
            			Left and right panels show the results of models M\_G2\_v400\_a05 and M\_G2\_v400\_a15, respectively.}
				\label{fig:drag}	
			\end{figure*}

       	 		To quantify the effect of the drag we compared the final binding energy of clumps in simulations with and without the presence of the drag force.
			In Figure~\ref{fig:drag}, we present histograms of the final binding energy of G2 candidates scaled by the orbital energy of IRS~16SW. 
        			Left and right panels present the outputs of models M\_G2\_v400\_a05 and M\_G2\_v400\_a15, respectively. 
        			Although we did not consider the presence of the drag force the power law of the ISM does affect the lifetime of clumps.
        			Notice that in both plots the presence of a drag force shifts the distribution towards larger binding energy ratios as we would expect (green and blue histograms). 
        			The drag acts subtracting kinetic energy making clumps to switch to more bound orbits, i.e., larger $E_f/E_i$ values. 
        			The effect is stronger when we consider a steep density profile ($\alpha=1.5$) as we observe a longer tail in the blue histogram of the right panel of Figure~\ref{fig:drag}. 
        			This makes sense as the density increases more rapidly which translates into a stronger hydrodynamical interaction between clumps and the medium.
        			In principle, this shows the drag effects are not negligible in our model. 
        			Nevertheless, despite it is significant we cannot distinguish whether or not they improve chances of reproducing G2 observations. 
        			Thus, the drag effect simply does not help clumps to mimic G2's motion close to Sgr~A*. 
        
		\subsubsection{Final clump masses}
     		\label{sec:mf}
     
\begin{figure}
				\centering
				\includegraphics[width=0.49\textwidth]{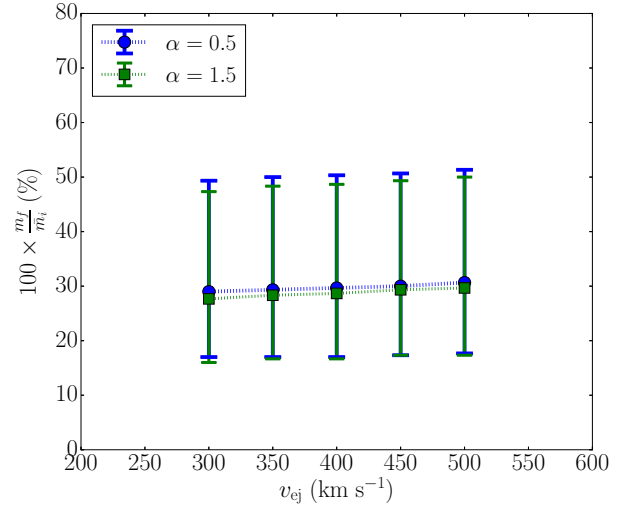}
				\caption{Median of the final clump mass distribution as a function of the ejection speed parameter. 
            			Error bars represent percentiles 34th and 68th. 
            			Final masses are expressed as percentages of the mean clump mass of the initial clump mass function which in these cases was set to $\bar{m}=3\rm\ M_{\oplus}$. 
            			Each point corresponds to a single simulation whose ejection speed is specified in the x-axis, and the density profile is color coded. 
            			Blue circles and green squares stand for the use of a shallow ($\alpha = 0.5$) and a steep ($\alpha = 1.5$) density profile, respectively. 
            			Notice there is no statistical difference among the choice of different model parameters.}
				\label{fig:mf}
			\end{figure}

     			As we have discussed clumps experience mass losses during their lives due to thermal conduction. 
        			Here, we refer as ``final mass" to the mass a clump would have at the present, i.e., $t=2016\rm\ yr$.
        			In Figure~\ref{fig:mf}, we show the final mass of simulations with $\bar{m_{\rm c}}=3\rm\ M_{\oplus}$ as a function of the ejection velocity $v_{\rm ej}$ of each run. 
			It is important to keep in mind that this analysis includes all clumps that have not been evaporated, and not only G2 candidates.
        			The density profile power-law is color coded being $\alpha=0.5$ and $\alpha = 1.5$, blue circles and green triangles, respectively. 
			Each point represents the median of the final masses while the error bars stand for percentiles 34th and 68th, respectively. 
        			In all models shown clumps had initial masses of $3^{+2}_{-1}\rm M_{\oplus}$ following a semi-log distribution. 
        			Notice that in all cases shown we see they have lost a significant fraction of their masses. 
        			Specifically, they have lost at least $50\%$ of their initial mass. 
        			Within the error bars both sets of simulations give exactly the same result. 
        			This means the choice of the ejection speed or the power-law of the density profile does not determine the amount of mass clumps lose. 
        			Instead this is more likely given by the period of time between the ejection and the end of the simulation.
        
        			Overall, clumps lose a significant fraction of their mass before being captured. 
        			Even the most massive clumps IRS~16SW can produce would lose $50\%$ of their mass. 
        			This means if a G2-mass clump was created by the binary, today we would observe it with at least half of its mass. 
        			Thus, G2's initial mass should have been at least $6\rm\ M_{\oplus}$. 
        			But our analytical estimates shows this is not feasible.
        			IRS~16SW is not capable of creating such massive clump given its wind properties.
     	
\section{Discussion}
\label{sec:disc}
    	 
	Our analytical and numerical results point that it is hard to conceive the origin of G2 in IRS~16SW, or in any other known massive binary.  
	Nevertheless, we have to be aware that there are some uncertainties our model did not take into account. 
	Here, we discuss the possible effect of such uncertainties.  
	Also, we comment on the detectability of gas clumps we expect IRS~16SW produces. 
       
	\subsection{Limitations and uncertainties in the model}
	\label{sec:lim}

		In our model we made two assumptions that could affect part of our results: the isotropic ejection of clumps, and not considering magnetic field effects. 
        
        		\subsubsection{Clump ejection isotropy}
        
			State-of-the-art numerical simulations of colliding wind binaries show that clumps are not ejected isotropically \citep{P09}. 
            		Instead, they are more likely launched perpendicularly to the binary orbital plane in some sort of cones upwards and downwards. 
            		Bear in mind that clumps are not uniquely ejected in these cones but preferably.
			Unfortunately, in the case of IRS~16SW it is not possible to estimate the inclination angle between the plane of the binary and the orbit around Sgr~A*. 
        			Thus, we just cannot constrain the direction of ejection. 
			However, in this scenario (in general) it is less likely for clumps to travel close to Sgr~A* compared to the isotropic case. 
			This is due to the fact that clumps need to be ejected on a given direction to get rid of most of their angular momentum, so they can fall into the SMBH.
			Therefore, if clump ejection occurs preferably on some sort of cones the chances of both directions to be aligned is smaller, the less isotropic the ejection takes place. 
			Based on this, the isotropy is the most sensible assumption to study this problem as it shows us the most optimistic situation.

		\subsubsection{Magnetic fields}
		\label{sec:mag}
        
        			Our approach does not include any effect of possible magnetic field present in the region. 
			As thermal conduction was considered in the saturation limit we do not expect the presence of a magnetic field affects it much. 
			However, the drag force could be modified by the action of magnetic fields. 
        			\cite{Mc15} showed that a magnetic field in a hot wind can enhance the drag force exerted on a cloud traveling through it. 
        			The increment is in a factor $(1+v_{\rm A}^2/v^2)$, where $v_{\rm A}=B/\sqrt{4\pi\rho}$ is the Alfv\'en speed and $v$ the relative speed between the cloud and the wind. 
        			Notice the strong dependence on the strength of the magnetic field. 
        			Thus, in this case the drag force has to be replaced by
        
        			\begin{equation}
        				F_{\rm drag}^{\rm mag} = F_{\rm drag}\left(1+\frac{2}{\beta\mathcal{M}^2}\right),
        			\end{equation}
        
        			\noindent where $\mathcal{M}$ is the Mach number, and $\beta=8\pi P_{\rm thermal}/B^2$ is the ratio of thermal to magnetic pressure of the wind.
        			Typically, at $r=1.4\rm\ arcsec$ the temperature of the medium is $10^7\rm\ K$ and clumps can reach speeds of $2000\rm\ km\ s^{-1}$, then $\mathcal{M}\sim 3$.
        			Therefore, in order for the drag to be enhanced by a factor two the beta parameter needs to be $\beta\sim 0.1$ (see Section~\ref{sec:fdrag}), i.e., a magnetic field of $B\sim10\rm\  mG$.  
        			In this region the only constraint on the magnetic field strength is given by the observations of the Galactic Centre magnetar \citep{E13}. 
			This suggests a magnetic field consistent with $\beta\sim1$ at $\sim0.1\rm\ pc$ ($2.5\rm\ arcsec$) from Sgr~A*.  
			Since we do not expect $B$ to depend strongly on $r$ based on most accretion models that assume equipartion \citep[see][]{E13}, the drag could be enhanced in a factor of two at most which is too small to have an impact on our results. 
        
	\subsection{Clump detectability}
			        
        		Regardless of whether G2 was created by IRS~16SW or not, we do expect clumps to be formed in the binary. 
		But, should we observe such clumps with our current observational facilities? 
		In this section we proceed to analyse if we could detect such objects. 
		In this way, we make sure that our description of IRS~16SW producing clumps is not in tension with observations.
        		To do so we study the Br-$\gamma$ luminosity a clump would radiate in this environment. 
        		We calculate the Br-$\gamma$ emissivity following case B recombination theory, i.e.,
		
        		\begin{equation}
        			j_{\rm Br\gamma}=3.44\times10^{-27}\left(\frac{T_{\rm c}}{10^4\rm\ K}\right)^{-1.09}n_{\rm p}n_{\rm e}\rm\ erg\ s^{-1}\ cm^3,
        		\end{equation}
        
        		\noindent where $n_{\rm p}$ and $n_{\rm e}$ are the proton and electron number density, respectively.  
		As discussed in Section~\ref{sec:ism}, it is reasonable to assume the temperature of the clump $T_{\rm c}$ to be $10^4\rm\ K$. 
        
        		In order to constrain the physical size of the clump in this environment we will assume it is in pressure equilibrium. 
		We are aware that a clump could be subject to tidal forces by the SMBH \citep[e.g.,][]{Ch16}. 
		However, we opted for studying the luminosity of the clump only at distances close to IRS~16SW. 
		At this location the thermal pressure of the environment dominates over the tidal forces on clumps whose masses are of the order of Earth's. 
		We estimated the role of the medium pressure and tidal force, and found out that at radius of $\sim1\rm\arcsec$ or shorter, the tidal force starts to dominate over the thermal pressure.		
        
\begin{figure}
			\centering
			\includegraphics[width=0.49\textwidth]{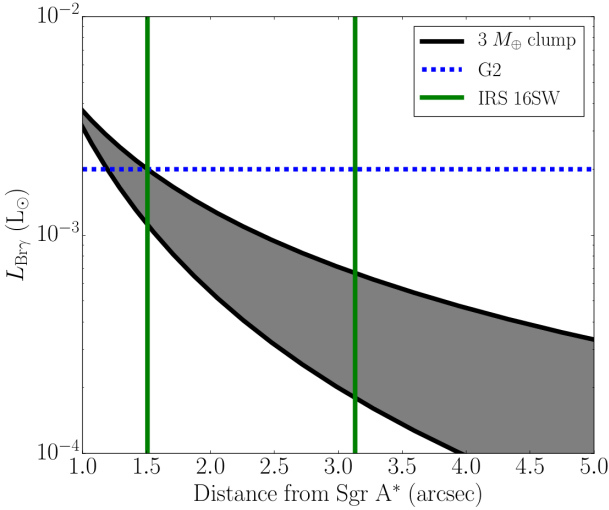}
			\caption{Br-$\gamma$ luminosity of a 3-$\rm M_{\oplus}$ clump as a function of distance from Sgr~A* (solid black line) estimated from our simple model. 
            		The observed Br-$\gamma$ luminosity of G2 is shown as a horizontal dashed blue line. 
            		Solid green vertical lines represent pericentre and apocentre of IRS~16SW around Sgr~A*}
			\label{fig:br_gamma}
		\end{figure}

        		In Figure~\ref{fig:br_gamma}, we show the Br-$\gamma$ luminosity $L_{\rm Br\gamma}$ of a 3 $\rm M_{\oplus}$ clump as a function of distance from Sgr~A*. 
		Solid black lines show the luminosity estimated when assuming a density power-law $a=0.5$ (lower limit) and $a=1.5$ (upper limit).
        		As comparison we show the Br-$\gamma$ luminosity of G2 constrained by observations to be $L_{\rm Br\gamma}=2\times10^{-3}L_{\odot}$ (dotted blue line). 
		
        		Vertical solid green lines stand for pericentre and apocentre distances of IRS~16SW. 
        		Notice that the maximum luminosity of clumps is on the pericentre of the binary, where it is very similar to G2's. 
		Further from this point the luminosity can only be of a fraction of G2's luminosity, so less likely to be detected. 
		If we consider less massive clumps they will be even fainter.  
		Then, only clumps that manage to travel towards Sgr~A*, and are about 3-$\rm M_{\oplus}$, could radiate as much as G2. 
		Unfortunately, we expect that most clumps ejected by IRS~16SW not to be detected. 
		The reasons behind this are that only few of them will be massive enough and, at the same time, launched in the direction of Sgr~A*. 
		Thus, our model and results are consistent with current observations.

\section{Conclusions}
\label{sec:conc}
    
        Our analytical and test-particle simulation results show that we cannot reconcile the hypothesis of G2 being a gas clump formed in any known massive binary system with its observed motion.  
        The results that support this conclusion are the following
        
        \begin{enumerate}
    
    		\item	\textit{G2 orbital fit is not consistent with an origin in IRS~16SW}.
        		It is not possible to match IRS~16SW and G2 positions on the sky if we trace their orbits back in time. 
        		Their projected separation (lower limit of the physical separation) always remains larger than $\sim0.2\rm\ arcsec$. 
        
        		\item	\textit{Cold gas clumps do not live long enough}. 
        		The hot environment close to Sgr~A* evaporates cold clumps very rapidly via thermal conduction. 
       	 	At the orbit of IRS~16SW, massive clump lifetimes and the free-fall timescale are comparable ($\sim200\rm\ yr$). 
        		Therefore, only the most massive of them would live long enough to reach the vicinity of Sgr~A*.
        
        		\item\textit{IRS~16SW cannot produce massive enough clumps}. 
		Given its orbital and stellar wind parameters, we can expect it to create clumps with {\it initial} masses of at most 3 Earth masses.
		Roughly half of that mass would be lost before reaching Sgr~A*, in disagreement with the observed G2 mass.
        
        		\item	\textit{Our test-particle simulations cannot reproduce G2 observations}. 
        		A model based on IRS~16SW ejecting clumps along its orbit is not capable of producing any clump that matches the observed G2 positions on the sky and radial velocity measurements. 
        
	\end{enumerate}
	
	These results, together with our previous work on encounters between single stars \citep{C16}, reject the idea of G2 being created in a stellar wind collision. 
	However, our work does not rule out the ``purely gaseous cloud" scenario in general. 
	Other gas cloud models remain as possible explanations. 
	In order to reject those, following a procedure similar to the one presented in this paper, it would be necessary to first identify candidate progenitors, e.g., a star recently going through a nova outburst, or being partially disrupted.
	  
\section*{Acknowledgments}

	We thank the anonymous referee for useful comments. 
	We also thank G.~Hajdu and A.~Ballone for very helpful discussions.
	This work was partially developed while JC was on sabbatical leave at MPE.  
	DC and JC acknowledge the kind hospitality of MPE, and funding from the Max Planck Society through a ``Partner Group'' grant.
	We acknowledge support from CONICYT-Chile through FONDECYT (1141175) and Basal (PFB0609) grants. 
	DC is supported by CONICYT-PCHA/Doctorado Nacional (2015-21151574).

\appendix
\section{Ambient medium model and clump sizes}	
\label{sec:ism}
	
	In order to estimate the mass loss through thermal conduction and to include the effects of the drag force in our simulations we need to specify density and temperature profiles for the environment. 
	In this work we consider the same approach used by \cite{B12}. 
	We used the model described by \cite{Y03} that reproduces \textit{Chandra} X-ray observations assuming a completely ionised gas and solar metallicity. 
        	This model is consistent with state-of-the-art modelling of these observations by \cite{R17}.
	Thus, the density profile of the medium is given by
	
	\begin{equation}
	\label{eq:dens}
		\rho_{\rm ISM}(r)=10^{-22}\left(\frac{1.7\times10^{17}{\rm cm}}{r}\right)^{\alpha}{\rm g\ cm^{-3}}.
	\end{equation}
			
	\noindent To compute a temperature profile we assume hydrostatic equilibrium between the hot ISM and the gravitational potential of Sgr~A* \citep{B12}. 
        	Therefore,

	\begin{equation}
	\label{eq:temp}
		T_{\rm ISM}(r)=\frac{2.4\times10^7}{\alpha+1}\left(\frac{1.7\times10^{17}{\rm cm}}{r}\right){\rm K}.
	\end{equation}
	
	With a thermodynamic description of the ISM we can estimate clump sizes. 
	To do so, we consider that clumps are completely ionised, and have a temperature of $T_{\rm c}=10^4{\rm K}$. 
	Both assumptions are justified because of the presence of strong UV field radiated by the massive stars in the inner parsec. 
	Also, we assume pressure equilibrium between the thermal pressure of the ISM and the clumps, i.e. $\rho_{\rm c}=\rho_{\rm ISM}T_{\rm ISM}/T_{\rm c}$. 
	Considering clumps with a given mass, uniform density and spherical symmetry, obtaining their radius is straightforward.

\label{lastpage}
\end{document}